# How human is the machine? Evidence from 66,000 Conversations with Large Language Models

Antonios Stamatogiannakis[a,1*], Arsham Ghodsinia[a,1], Sepehr Etminanrad [b,2], Dilney Gonçalves [a,2], David Santos [c,2]

[a] IE Business School, IE University; Madrid, Spain; 28006.

[b] Rotterdam School of Management, Erasmus University; Rotterdam, Netherlands; 3062 PA.

[c] Psychology Department, Universidad Autónoma de Madrid; Madrid, Spain; 28049.

* Corresponding Author: Antonios Stamatogiannakis**,** email:  antonios.stamatogiannakis@ie.edu

**Author Contributions:** The first two authors contributed equally to this work and should be considered as equal first co-authors.

**Keywords:** AI, GPT, heuristics, decision-making, algorithms, large language models.

## Abstract

When Artificial Intelligence (AI) is used to replace consumers (e.g., synthetic data), it is often assumed that AI emulates established consumers, and more generally human behaviors. Ten experiments with Large Language Models (LLMs) investigate if this is true in the domain of well-documented biases and heuristics. Across studies we observe four distinct types of deviations from human-like behavior. First, in some cases, LLMs reduce or correct biases observed in humans. Second, in other cases, LLMs amplify these same biases. Third, and perhaps most intriguingly, LLMs sometimes exhibit biases opposite to those found in humans. Fourth, LLMs' responses to the same (or similar) prompts tend to be inconsistent (a) within the same model after a time delay, (b) across models, and (c) among independent research studies. Such inconsistencies can be uncharacteristic of humans and suggest that, at least at one point, LLMs' responses differed from humans. Overall, unhuman-like responses are problematic when LLMs are used to mimic or predict consumer behavior. These findings complement research on synthetic consumer data by showing that sources of bias are not necessarily human-centric. They also contribute to the debate about the tasks for which consumers, and more generally humans, can be replaced by AI.

## Introduction

Consumers often use AI for diverse set of tasks and as Artificial Intelligence (AI) and Large Language Models (LLMs) evolve, one of their key uses is to replace consumers, and more generally humans in domains such as communication (Xiao & Yu, 2025), medicine (Haim et al., 2024), businesses and organizations (Dell'Acqua et al., 2023), and algorithmic learning (Lee et al., 2023). These cases of replacement are supported by emerging academic and business research, often showcasing the superior performance that newer LLMs often exhibit in comparison to older ones (Kosinski, 2024) and to humans (Gilardi et al., 2023).

Whereas the above literature thoroughly studies LLMs' performance in various domains, a different ability that generative AI in general and LLMs in particular are assumed to have when replacing consumers, and more generally is that they are representative of human behavior (Dillion et al., 2023; Hermann & Puntoni, 2024; Yax et al., 2024). Accurately capturing consumer behavior is different from, and often incompatible with, improving performance. For instance, to capture consumer behavior, AI should not correct consumer shortcomings, but it should replicate them in terms of magnitude and direction.

This ability, which we call Human-likeness, has received scant research attention and mixed empirical support. For instance, although acting like humans seemed to be initially an undisputed property of LLMs (Acerbi & Stubbersfield, 2023; Hagendorff et al., 2023; Suri et al., 2024), controlled experiments showed that popular LLMs resembled human behavior only in about 50% of the cases (Chen et al., 2025). The present manuscript contributes to research about the human-likeness of LLMs by demonstrating different ways in which an LLM can deviate from human-likeness in classic judgment and decision-making problems. We show deviations from human-likeness in the form of 1) attenuating or fixing human biases, 2) exaggerating human biases, 3) reversing human biases, or 4) producing inconsistent responses in cases where consumers, and more generally show consistency.

We study these deviations from human-likeness in six important decision-making biases and heuristics. Three allow for objectively correct responses: availability and representativeness heuristics, and the endowment effect (Kahneman et al., 1990; Tversky & Kahneman, 1973, 1974)[1]. For the other three, there is no objectively correct response, but it is well-established how contextual factors are influential: anchoring, transaction utility, and framing (Thaler, 1985; Tversky & Kahneman, 1973, 1974, 1981, 1983). Our empirical investigations comprise ten pre-registered experiments, involving 66,000 conversations with large language models. Across these conversations, we systematically vary both stable characteristics of the LLM (i.e., model), as well as aspects of the prompts used.

Testing an extensive spectrum of deviations from human-likeness contributes to theory, practice, and policy. Theoretically, we contribute by studying a relatively neglected, but important consideration of replacing consumers, and more generally humans with AI: whether AI behavior can systematically deviate from consumers, and more generally human behavior, and cases where this deviation makes replacement problematic. Importantly, testing whether LLMs can have their own, unique responses, which are different, but not necessarily better than those of humans, can help us understand the extent to which AI behavior theories should assimilate psychological theories of consumers, and more generally human behavior or evolve independently.

Our results also have important practical implications. Four hundred million people use ChatGPT every week (Rooney, 2025), and major organizations such as IBM are replacing humans with AI for purposes such as human resources (Williams, 2025). To these users and organizations, we suggest caution for either using AI to replace human labor or for providing AI-based services as an alternative to human-based ones to their customers (e.g., replacing customer care agents with chatbots or using synthetic consumer data for market research purposes).

---

[1] Our design allows for objectively correct responses for the endowment effect because we use a variation in which market price is known

Next, we review the literature on the extent to which AI can replace consumers, and more generally humans, the existing evidence that directly or indirectly supports the human-likeness, as well as the evidence that argues against it in four possible forms: bias reduction, amplification, reversal, and inconsistencies over time.

## Conceptual Background

*Replacement of humans by AI*

The replacement of humans by AI has been argued to provide numerous benefits, such as higher efficiency (Dell'Acqua et al., 2023; Castelo, 2024), performance (Gilardi et al., 2023; Luo et al., 2024; Schoenegger et al., 2024), or flexibility (Dillion et al., 2023; Tomaino et al., 2025). Indeed, since the appearance of LLMs, human labor has been less demanded for writing, coding, and design tasks (Demirci et al., 2025).

Despite these benefits, the replacement of humans by AI often assumes that the latter's behavior resembles that of the former. Such human-likeness of AI behavior differs from the performance and efficiency benefits mentioned above. For example, to predict or mimic consumer responses in market research, AI should not correct or change sub-optimal choices or mis-calibrated preferences. Instead, it should err similarly to consumers in terms of magnitude and direction, and demonstrate consistency over time, like humans do (Cialdini, 2001). So, do LLMs behave like humans? We examine this question next.

*Similarities between AI and human behavior*

A common tenet behind the development of AI is that it aims to emulate consumer behavior (Li et al., 2024). For instance, LLMs are customized to emulate human psyche/personality profiles (Wang et al., 2025). The expectation of human-likeness is underlying the frequent evaluation of LLMs by comparing their behavior to human behavior in tasks like medical diagnoses (Feng et al., 2025), abstract and analogical reasoning (Webb et al., 2023), and technical tasks such as code generation (Ziems et al., 2023). This expectation is also implicit in the business practice of replacing humans with AI agents in organizations (Tabrizi & Pahlavan, 2023) or in testing with AI psychological concepts such as the theory of mind (Strachan et al., 2024; Kosinski, 2024).

In some respects, previous research supports the idea that LLMs can behave like humans. For example, in a Turing test (1950), individuals could not distinguish GPT-4 from human conversation partners (Jones & Bergen, 2024). Similarly, LLMs can be used to derive satisfactory assessments of consumer preferences based on well-defined attributes (Li et al., 2024), to achieve human-like creativity (Castelo et al., 2024), to personalize messages to customers (Matz et al., 2024), and so on.

Even when AI models err, their errors are often attributed to sources related to human limitations and hence are expected to be analogous to human errors. There are documented cases of automated AI biases ranging from semantic association structures (Caliskan et al., 2017) to online advertising (Sweeney, 2013) and criminal sentencing (Angwin et al., 2022). Even LLMs that can understand and generate human-like language (OpenAI, 2023) can be convinced with human persuasion techniques to provide harmful answers, even if they are explicitly coded not to (Zeng et al., 2024).

The cause of these errors is often assumed to be either the human-generated dataset based on which algorithms are trained (Morewedge et al., 2023) or the human-designed training process in which human biases may penetrate undetected (Latif & Zhai, 2024; Sweeney, 2013). Despite these similarities, LLMs' similarity to humans in the domain of heuristics and biases has been argued to be about 50% (Chen et al., 2025).

We build on this research by introducing a framework of the different ways in which LLM behavior can deviate from that of consumers, and more generally humans and then testing how human-like the responses of popular LLMs are, based on this framework.

*Differences between AI and human behavior*

AI can have important differences from consumers, and more generally humans that may lead to deviations in behavior. In line with recent calls for more nuanced, non-binary results (McShane et al., 2024), the first two deviations pertain to whether LLMs partially or completely fix human biases or, contrarily, amplify them. First, LLMs are trained and updated using enormous computational power and are not influenced by fatigue. Thus, they may be less prone to biases that often stem from cognitive limitations (Gilovich et al., 2002). Hence, the most important evidence for deviations from human-likeness is evidence for performance improvements (Gilardi et al., 2023; Luo et al., 2024; Schoenegger et al., 2024).

Besides improving human performance, a second way LLMs may behave differently than humans is to exaggerate their errors and biases. Such a result would likely emerge from the joint operation of two processes. First, the abundance of human errors and biases in the training datasets of LLMs. Second, hyper-accuracy distortion, a process in which LLMs' responses are a hyper-accurate representation of their training data (Aher et al., 2023). If an LLM is asked, for example, to behave like research respondents, these two processes could result in LLM behavior that over-mimics consumers, and more generally humans, in the form of presenting errors and biases of the same direction, but of larger magnitudes than humans.

A third way LLMs may behave differently than humans would be to produce biases or errors in the opposite direction. This can be, for example, the case of an LLM over-correcting human biases. As another example, humans tend to make many more mistakes when performing advanced, versus simple, mathematical calculations. On the contrary, LLMs are capable of advanced mathematical calculations, but often struggle with basic counting (Xu & Ma, 2024) or mathematical operations (Chen et al., 2023). Hence, tests that allow LLMs to err differently than humans are crucial to identify potential LLM weaknesses that differ from those of humans.

Following the reasoning above, the heuristics and biases that characterize human cognition and decision making (Gilovich et al., 2002) are an appropriate domain to properly test and understand the human-likeness of LLM behavior. Some of them (e.g., availability and representativeness heuristics) allow for objectively correct answers, against which the answers of LLMs can be compared to test if they mimic, resolve, or exaggerate human biases. In others (e.g., transaction utility and framing), there is no objectively correct response, but it is well-established how potentially irrelevant contextual factors influence behavior.

Hence, besides testing if LLMs mimic, resolve, or exaggerate consumers, and more generally human biases, these domains also allow for testing if LLMs' responses exhibit their own, unique biases (showing a reversal). For example, in the well-known anchoring effect (Tversky & Kahneman, 1974), human responses tend to be higher (vs. lower) when respondents are provided with an arbitrarily high (vs. low) anchor before the judgment. Would LLMs show the same bias, would their responses be unaffected by the anchors, or could they show a reverse bias? Some recent evidence on this topic indicates that early LLM models are likely to exhibit biases similar to humans (Chen et al., 2025; Hagendorff et al., 2023; Suri et al., 2024; Yax et al., 2024)[2].

Our results do not confirm this finding, showing the value of systematically testing different types of deviations from human-likeness. Hence, we contribute to this research by thoroughly testing LLMs' human-likeness: Whether LLMs tend to fix, exaggerate, or reverse human biases, producing their own unique biases. We further test whether this (dis)-similarity may be influenced by LLM characteristics (i.e., model advancement; GPT-3.5-turbo vs. GPT-4 across studies; 10 different GPT models in study 7) as well as by the prompts provided by users.

---

[2] Although in general aligned with our approach, these investigations have methodological limitations such as the excessive reliance on early LLMs, small sample sizes, ad-hoc experimental stimuli, no pre-registrations, use of the chat version of the LLM (please refer to the methodology section for details about the limitations of the chat version), and lack of human role assignments that can by itself explain deviations from human-likeness. Hence, we treat these investigations as initial tests of the human-likeness. Largely, this initial evidence does not include controlled comparisons between different LLMs, as our work.

*Consistency of LLM behavior as an indication of human-likeness*

Consistency of LLM behavior relates to human-likeness in two distinct ways. First, consistency is thought to be characteristic of human behavior (Cialdini, 2001; Festinger, 1957). For instance, past behavior is the best predictor of future behavior (Ouellette & Wood, 1998; Ozer & Benet-Martínez, 2006), and consumers consistently engage in behaviors they find enjoyable or useful, even more than they themselves predict (Kahneman & Snell, 1992).

Furthermore, individuals are asked to indicate future behaviors (e.g., in political polls or market research), assuming that these can inform decisions months later (Nordin & Ravald, 2023). Finally, in academic research, important findings such as the endowment effect (Kahneman, Knetsch, & Thaler, 1990) consistently replicate even decades later (Morewedge et al., 2009; Reb & Connolly, 2007). Hence, if AI replaces humans, it should also exhibit consistent responses over time.

Second, besides being characteristic of consumers, and more generally human behavior, consistency is also a necessary condition for human-likeness: If AI behavior changes, then at least at some points in time it is likely to be different from that of humans. Based on extant understanding, LLM behavior may fluctuate as a function of (a) their training dataset (Morewedge et al., 2023), (b) their training process (Latif & Zhai, 2024), and (c) the prompt used (Meincke et al., 2025). Hence, one would expect the exact same LLM, with the same training dataset and process, to provide similar answers to the same prompt (after allowing for random variance), even with a time delay. We test this prediction.

The behavior of successive LLMs within the same family, though, could differ. For example, even LLMs from the same family have shown diverse political stances (Rozado, 2024). Similarly, newer versions demonstrate improvements in language understanding, generation, contextual awareness, and cognitive tasks (Yax et al., 2024). Such constant improvements, however, imply that early LLMs were worse (hence different) than humans, and future ones may be better (and hence, again, may be different) than humans.

In summary, although the same model may behave consistently under the same conditions, different versions of LLMs may behave differently. Newer (vs. older) models may not only exhibit differences in performance but also unexpected qualitatively different behaviors. Besides being atypical of consumers, and more generally human behavior, such inconsistencies also indirectly indicate deviations from human-likeness: Inconsistent LLM behavior across versions means that at least some versions behave differently than humans. Hence, we test the extent to which successive LLM versions behave consistently.

Of course, inconsistent behaviors across models may arise if more advanced models can follow instructions to behave like humans better than older models. However, preliminary evidence suggests that this is questionable, with older models following instructions better than newer ones under some conditions. For example, Chen et al. (2023) found that the less capable GPT-3.5 can outperform the more capable GPT-4 in mathematical questions. Although this finding can be attributed to fine-tuning (Latif & Zhai, 2024) or to the precise prompt used (Henrickson & Meroño-Peñuela, 2023), it raises the possibility that, under some conditions, older models may be able to behave more human-like, if asked to do so. To test this possibility, we instruct LLMs to behave like human research participants in at least some conditions in all our studies.

*Scope of Empirical Investigation*

To test the human-likeness, we examine how LLMs respond to some of the most established tests of human biases and heuristics and compare these responses with the typical ones found for human decision-making. To study the consistency of LLM responses, we compare the responses of two LLMs (GPT-3.5-turbo and GPT-4) that belong to the same family. Although more advanced models are currently available, our interest is not in any model per se; rather, we use them to investigate (a) potential inconsistencies between the responses of different models, hinting at potential deviations from human-likeness, and (b) whether model advancement alone can be expected to mitigate such deviations. We complement these pairwise comparisons with a final study comparing 10 GPT models, including those using advanced reasoning.

Across studies, we further test the response consistency of LLMs, as well as the response consistency between our results and those of research using similar methodology. Finally, we test LLMs' variance of responses and differences in responses as a function of prompt variations, and discuss how these findings connect with human-likeness.

We acknowledge upfront that our research does not aim to, and hence is not designed to, uncover the theoretical processes underlying the effects we observe, as this task is likely beyond the scope of a single paper. For example, research on the human heuristics we study evolved for decades around underlying processes (e.g., Tversky & Kahneman, 1973; Vaughn, 1999). Nevertheless, in the general discussion section, we discuss potential processes underlying our findings.

## Methodology of Empirical Investigations

### *Experimental Procedures*

All our empirical studies shared the same methodology and followed open science guidelines (pre-registered, open materials, open data and code; with detailed explanation, including the Methodological Detail Appendix (MDA), are available in the OSF repository). We collected observations using our own custom-built application[3] written in Node.js. This application sent requests to the Application Programming Interface (API) of OpenAI, the company behind ChatGPT among other AI models. Overall, this method is comparable to recruiting human participants for an experimental study. The observations are independent (in each iteration—equivalent to a human participant—the model has no knowledge of other iterations), and each "participant" is randomly assigned to only one of the experimental conditions.

Using the API to prompt the models instead of accessing them through the web version has several advantages. First, using the API enables researchers to compare research with similar methodologies but with different "temperatures." Temperature is a parameter of LLM that can take values between 0 and 2 (OpenAI, 2023), with higher temperature leading to more creative responses and lower temperature to more deterministic ones. We used the default temperature of "1," balancing between determinism and creativity. Second, the API gives greater control over the precise version of a model used. Models often remain accessible even after updates, unlike the Chat version, whose model is irreversibly updated.

Third, this method provided operational efficiency to sample thousands of responses within a few hours and hence reduce potential confounds of time (e.g., GPT model updates). To be more specific, the data collection for each of our studies was completed in approximately 2 to 5 hours. Relatedly, conversations in the API platform are not used for model training, controlling for the possibility that later conversations would be influenced by training based on the former.

Fourth, this approach enabled us to study LLMs simultaneously, controlling for the influence of time (Chen et al., 2023). Fifth, this method reduces potential data entry errors by automatically creating time-stamped datasets. Sixth, numerous companies use the same API as our studies, including most Fortune 500 companies (Rooney, 2025). An example is the Microsoft Azure cloud computing platform for businesses (Microsoft, 2024).

---

[3] Access to this application is available upon request to the authors. We plan to make it completely public and free to access, bundled with a methodology paper explaining its use.

In each trial[4], corresponding to an observation in our datasets, the LLM was presented with a study designed to replicate closely one of the heuristics or biases studied: the availability heuristic (Tversky & Kahneman, 1983), the representativeness heuristic (Tversky & Kahneman, 1974), the endowment effect (Kahneman et al., 1990), the anchoring and adjustment heuristic (Tversky & Kahneman, 1974), the transaction utility effect (Thaler, 1985), and the framing effect (Tversky & Kahneman, 1973). For studies 1–6.3, we collected 1,000 observations for each condition (see MDA for the experimental designs). For study 7, we collected 100 observations per condition, enough to detect effects based on study 3.

Six studies (studies 1, 2, 3, 4, 5.1, and 6.1) were conducted in September and October 2023, three more (5.2, 6.2, and 6.3) in May 2024, and study 7 was conducted in December 2024.

For the first set of studies, each response was obtained with a sequence of two prompts. In the first prompt, following best practices (Chen et al., 2023; OpenAI, 2023), we provided some context and asked the LLM to assume the role of a research participant (see the MDA for the exact prompt). This was the role that humans had in the studies that established the findings in the domains we investigate; hence, this role assignment is a conservative prompt, controlling for the possibility that LLMs behave unlike humans simply because they were not explicitly asked to emulate their behavior.

We manipulated the presence of human role assignment in studies 5.2, 6.2, and 6.3, including a 7-point rating scale manipulation check to assess the role that the LLM perceived it was following. Hence, these studies systematically checked for the influence of role assignment. The second prompt asked about the question of interest in each study. Like surveys that ask human participants to respond in a specific way (e.g., on rating scales), and consistent with guidelines (OpenAI, 2023), the GPT model was requested to answer in a structured format.

---

[4] There were minor modifications in what is described in this section in some studies. When this happens, we explain these modifications in detail in the MDA.

In each study, we added an additional 2 × 2 experimental design to the core effect under investigation. The first factor of this design was the GPT model version. We studied each effect on two widely used LLMs, GPT-3.5 and GPT-4, with the latter considered superior to the former (OpenAI, 2023).

The second factor was whether the LLM was provided with a short induction, which could improve response accuracy (Li et al., 2024). We used a two-sentence debiasing instruction (vs. no debiasing) for half of the trials in each study to test whether such minimal instruction could induce LLMs to correct for biases (see the MDA). No such instruction was given for the other half of the trials (no debiasing condition). This manipulation was largely exploratory. For instance, it could have little influence if the LLM cannot detect a bias.

Studies 5.2, 6.2, and 6.3 had a complementary design. Instead of the debiasing instructions, we manipulated the presence of the human role assignment prompt. Additionally, the data for these studies were collected eight months later than the first six studies, allowing us to investigate the effect of time in comparable conditions similar to Chen et al. (2023).

Finally, study 7 used the procedure of the first studies with only LLM type as a manipulated factor. We compared the responses of 10 different GPT LLMs, including the most powerful ones at the time of the study, to the well-known endowment effect. This study helped us examine different deviations from human-likeness more thoroughly, even among the most recent LLMs.

*Identification of Bias Attenuation, Amplification, and Reversal*

Our research aimed to compare the behavior of LLMs to that of humans and assess whether they are similar, or LLMs attenuate, amplify, or reverse human biases. In this section, we establish the procedure through which we identified bias attenuation, amplification, and reversal.

We started by performing the appropriate statistical tests to identify direction, significance levels, and sizes for the effects in each of our studies, according to our pre-registrations. Then, we sought the literature to identify meta-analyses about the domains we studied: availability heuristic, representativeness heuristic, endowment effect, anchoring, transaction utility, and framing. We were able to detect such meta-analyses for three of them: endowment meta-analysis by Sayman and Öncüler (2005), anchoring by Schley and Weingarten (2025), framing by Steiger and Kühberger (2018), and Piñon and Gambara (2005). As these meta-analyses followed different procedures, we identified in each the core measurement on which the meta-analysis of the effect was performed.

Subsequently, for each of these domains, we compared our results to those of the meta-analysis as follows. We marked as "attenuations" cases where the effect in our studies (a) was not significant or (b) was significant and of the same direction, but weaker than the lowest point of the 95% confidence interval of the reported effect size, or, if no confidence interval was reported, our effect value (e.g., mean, ratio, etc.) was substantially below the reported one in the meta-analysis. We marked as "amplification" cases where the effect in our studies was of the same direction, but stronger than the highest point of the 95% confidence interval of the reported effect size, or, if no confidence interval was reported, our effect value (e.g., mean, ratio, etc.) was substantially below the reported one in the meta-analysis. We marked as "reversals" cases where the effect in our studies (a) was significant and (b) of the opposite direction than the one reported in the meta-analysis. Regardless, we reported relative differences in percentage[5] for all comparisons (our vs. the original/meta-analysis findings) in Table MDA1.

We could not detect such meta-analyses for the other three domains we studied. For these domains, we based our comparisons on the results of the core measurement reported in the original research of each: availability heuristic (Tversky & Kahneman, 1973), representativeness heuristic (Tversky & Kahneman, 1983), and transaction utility (Thaler, 1985). This was an appropriate comparison standard because our experimental procedures were almost identical to these classic works. We marked as "attenuations" cases where the effect in our studies (a) was not significant or (b) was significant and of the same direction, but substantially weaker (for the exact numerical values see MDA's Table MDA1) than the original finding. We marked as "amplification" cases where the effect in our studies was of the same direction, but substantially stronger than the original finding. We marked as "reversals" cases where the effect in our studies (a) was significant and (b) of the opposite direction than the original finding[6].

---

[5] Relative difference in percentage was calculated as: $\Delta\% = (\text{our findings} - \text{baseline}) * 100/\text{baseline}$

Through our studies, we observed multiple significant interaction effects, indicating that the core effect under investigation was moderated by factors such as the GPT model, the presence of debiasing prompts, or human role assignment. Hence, we performed and reported the aforementioned comparisons separately for each experimental condition. These comparisons are summarized in Table 1, and in greater detail in the MDA's Table MDA1.

**Results**

We now summarize the main results obtained with the above methodology. The studies are ordered beginning with scenarios that have an objective answer, followed by those that do not. Study 7 is presented last because it has a different design (sample size and factors) and was conducted at a later time. The main conclusions about how LLMs deviated from human-likeness in all 10 studies appear in Table 1. Results are visualized in Figure 1. Effect sizes appear in Table 2. Complete descriptive, inferential and variance tests results are available in the MDA.  Table 1 shows that in none of the cases, even when LLMs were instructed to behave like research

---

[6] As a robustness check, we compared our results with both existing meta-analyses, and the originally reported effects, for the endowment and framing domains that this was possible. The outcomes of these two comparisons were the same for 25/26 conditions involved in these studies.

participants, we observed a bias similar in direction and magnitude to established human biases.

We now elaborate on each of the identified deviations.

**Table 1.** *Deviations from human-likeness for each model, study, and condition.*

| *GPT Model* | GPT-3.5 | | GPT-4 | |
|---|---|---|---|---|
| *Debiasing /Role Instructions* | Research Participant Role | Research Participant Role with Debiasing (studies 1, 2, 3, 4, 5.1, 6.1) OR No Role (studies 5.2, 6.2, 6.3) | Research Participant Role | Research Participant Role with Debiasing (studies 1, 2, 3, 4, 5.1, 6.1) OR No Role (studies 5.2, 6.2, 6.3) |
| *Study 1 Availability* | Attenuated Bias | Attenuated Bias | Amplified Bias | Amplified Bias |
| *Study 2 Representativeness* | Attenuated Bias | Attenuated Bias | Amplified Bias | Amplified Bias |
| *Study 3 Endowment* | *Reverse Bias* | *Reverse Bias* | Attenuated Bias | Attenuated Bias |
| *Study 4 Anchoring* | Amplified Bias | Amplified Bias | *Reverse Bias* | Attenuated Bias |
| *Study 5.1 Transaction Utility* | Attenuated Bias | Attenuated Bias | Attenuated Bias | Attenuated Bias |
| *Study 5.2 Transaction Utility x Role* | Attenuated Bias | Attenuated Bias | Attenuated Bias | Attenuated Bias |
| *Study 6.1 Framing Lives* | Attenuated Bias Riskier than humans | Attenuated Bias Riskier than humans | Attenuated Bias Less risky than humans | Amplified Bias Less risky than humans |
| *Study 6.2 Framing Lives x Role* | Attenuated Bias Riskier than humans | Amplified Bias Riskier than humans | Attenuated Bias Less risky than humans | *Reverse Bias* Less risky than humans |
| *Study 6.3 Framing Jobs x Role* | Attenuated Bias Riskier than humans | Attenuated Bias Riskier than humans | Amplified Bias Less risky than humans | *Reverse Bias* Riskier than humans |
| *Study 7 Endowment Across Models* | 9 cases of bias attenuation (Gpt-3.5-turbo-0125, Gpt-3.5-turbo-1106, Gpt-4-0125-preview, Gpt-4-0613, Gpt-4-turbo-2024-04-09, Gpt-4o-2024-08-06, Chatgpt-4o-latest, O1-mini-2024-09-12, and O1-preview-2024-09-12) and 1 case of reversal (Gpt-4o-mini-2024-07-18) | | | |

Note: The prompt in studies 1, 2, 3, 4, 5.1, 6.1 always assigned the role of a research participant, and had a debiasing manipulation. The prompt in studies that have "x Role" in their title (studies 5.2, 6.2, and 6.3) manipulated research participant role vs. no role. Study 7 did not have a prompt-based manipulation.

**Figure 1, Panel A.** *Results[7] for all studies, with 95% Confidence Intervals. Except for studies 1, 2, and 7 (different design), the left panels represent GPT-3.5, and the right ones GPT-4 for all other studies.*

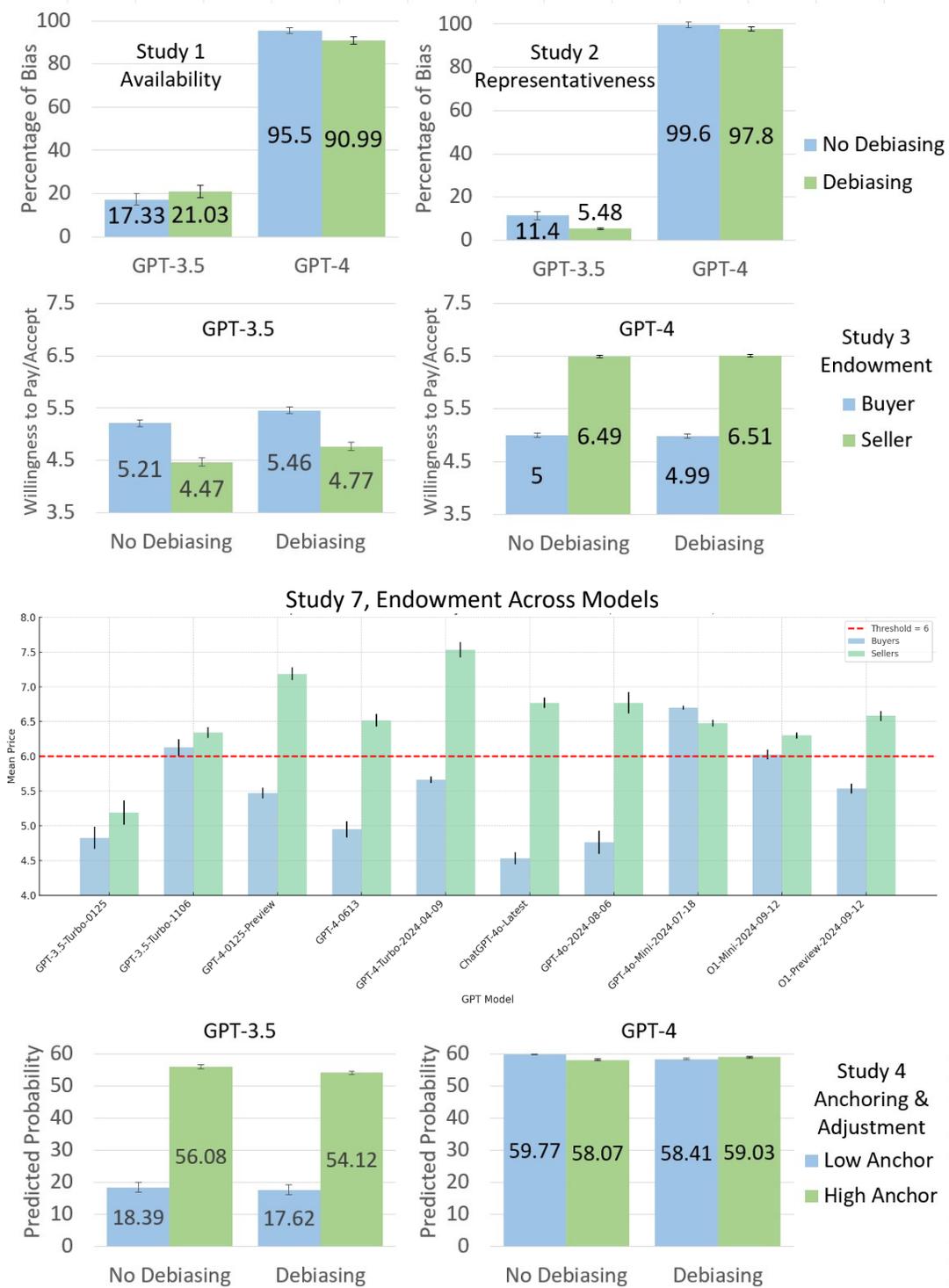

**Figure 1, Panel B.** *Results for all studies, with 95% Confidence Intervals. The left panels are for GPT-3.5, and the right panels are for GPT-4*

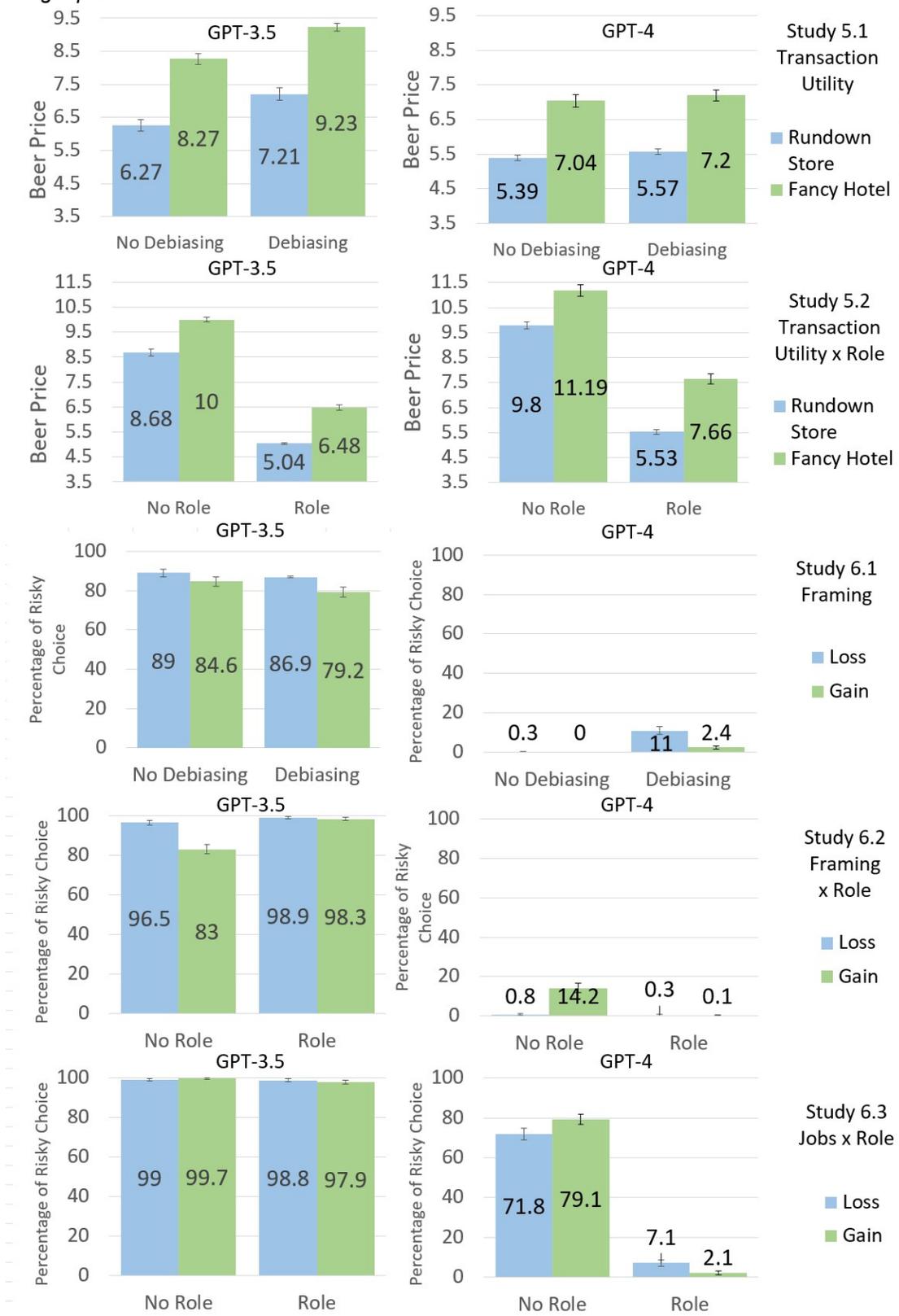

**Table 2.** *Summary of the statistical models' results (effect sizes –partial $\eta^2$ for Analyses of variance (ANOVAs) and odds ratio for logistic regressions– and significance levels).*

| Study (Method) | Focal Manipulation (a) | GPT Model (b) | Debiasing / Role (c) | a × b | a × c | b × c | a × b × c |
|---|---|---|---|---|---|---|---|
| Study 1 Availability (logistic regression) | NA | 7.87*** | 0.88* | NA | NA | 0.78*** | NA |
| Study 2: Representativeness (logistic regression) | NA | 34.89*** | 0.53*** | NA | NA | NS | NA |
| Study 3 Endowment (ANOVA) | 0.05*** | 0.17*** | 0.01*** | 0.3*** | NS | 0.01*** | NS |
| Study 4 Anchoring (ANOVA) | 0.321*** | 0.412*** | 0.001** | 0.334*** | NS | < 0.001† | 0.001** |
| Study 5.1 Transaction Utility (ANOVA) | 0.13*** | 0.09*** | 0.01*** | 0.002*** | NS | 0.01*** | NS |
| Study 5.2 Transaction Utility x Role (ANOVA) | 0.108*** | 0.046*** | 0.406*** | 0.002*** | 0.002*** | 0.001** | 0.001** |
| Study 6.1 Framing (Lives) (logistic regression) | 0.79*** | NA | 0.87*** | NA | NS | NA | NA |
| Study 6.2 Framing (Lives) x Role (logistic regression) | NS | 0.02*** | 0.74† | 1.67** | 0.70* | 0.30*** | 0.51*** |
| Study 6.3 Framing (Jobs) x Role (logistic regression) | NS | 0.06*** | 0.26*** | 0.83† | 0.65*** | 0.44*** | NS |
| Study 7 Endowment across models (ANOVA) | 0.53*** | 0.439*** | NA | 0.395*** | NA | NA | NA |

***$p < .001$, **$p < .01$, *$p < .05$, †$p < .1$, NA = not applicable because of study design, NS = not significant. The analysis for Study 6.1 is only about GPT-3.5 because of quasi-separation for GPT-4. Study 7 had a 2x10 design and 100 observations per cell (instead of 1,000 as in other studies).

*Bias attenuation*

Bias attenuation was the most frequent deviation from human-likeness, observed in each of the 10 studies in at least one condition (Table 1). On the one hand, this finding is consistent with research suggesting that LLMs can outperform humans, which in our case translates to exhibiting fewer or less severe biases. This bias attenuation, however, was mostly partial, as in only 6 out of 31 bias attenuation cases, the LLMs completely fixed the human bias (non-significant differences across key conditions). On the other hand, it is problematic that bias attenuation was very frequent even in conditions in which LLMs (GPT-3.5: all studies except study 3 and 4; GPT-4: studies 3, 5.1-5.2, 6.1-6.3, 7; seven additional models in study 7) were instructed to behave like a research participant, and hence reproduce human biases. The extensive bias attenuation pattern suggests that this may be the most important breach of human-likeness in LLM behavior.

*Bias amplification*

Bias amplification was observed for GPT-3.5 in study 4 (anchoring) and 6.2 (framing x role) and for GPT-4, in four studies (studies 1, 2, 6.1 and 6.3). For instance, in the case of GPT-3.5 in study 4, the observed effect sizes for both debiasing and no debiasing conditions were considerably higher than the human-like anchoring effects reported in a meta-analysis of 2,603 anchoring effect sizes (Schley & Weingarten, 2025).

As for GPT-4, interestingly, studies 1 and 2 both had objectively correct responses, so the fact that GPT-4 was biased in more than 90% of the cases poses another threat to human-likeness. Of note, the fact that bias amplification was much more rare than bias attenuation (nine in comparison to 31 conditions, across studies) runs counter to the possibility that LLMs behave on average like humans, by sometimes attenuating, and others amplifying human biases. Rather, it seems that these behaviors may be driven by distinct bias-specific processes.

Relatedly, although not a bias, GPT-3.5 showed amplified risky behavior in comparison to humans to a large degree: For the "gain frame" conditions of the framing studies 6.1-6.3, risky choices varied between 79.2% and 99.7%, compared to 28% originally reported for these conditions (Tversky & Kahneman, 1981). Similarly, for the "losses" conditions in these studies, risky choices varied between 86.9% and 99%, compared to 78% originally reported for these conditions (Tversky & Kahneman, 1981).

*Reverse biases*

The most intriguing deviation from human-likeness is perhaps reverse biases, which we observed in some conditions in five of our studies. The occurrence of these reverse biases does not seem to be systematic. In all the studies in which we observed them, we also observed bias attenuation in some conditions. Also, we observed reverse biases (a) with three different models (GPT-3.5, GPT-4, and GPT-4o-mini), and (b) both when LLMs were taking the role of a research participant (Studies 3, 4, and 7), and when they had no role (Studies 6.2 and 6.3).

*Consistency of responses*

Finally, we provide detailed results for three types of response consistency in the MDA. Within-model consistency results show that the exact same model provided significantly different responses to the exact same prompt after time delay, in both[8] occasions allowing for such analysis (Studies 5.1 vs. 5.2, and 6.1 vs. 6.2). Cross-model consistency results show that the "GPT model" factor also resulted in both significant main and interaction effects (with other independent factors in each study). In eight out of the nine studies that included prompt variations we observed a significant "prompt variation X model" interaction, illustrating that the two models handled minor prompt variations differently.

Finally, cross-research consistency results show that researchers using similar methodologies (our work; Chen et al., 2025; Suri et al., 2024) produce inconsistent results. Suri et al. (2024) find that LLMs mimic human responses. In contrast, for the same domains of study, we found that at least in some conditions, LLM responses are different from those of humans in the form of either bias attenuation or bias reversal. Additionally, the results of Chen et al. (2025) for the representativeness heuristic (conjunction fallacy), the availability heuristic, and framing are, on average, different from the results of both Suri et al. (2024) and our results. For the endowment effect, the results were more aligned, although there were deviations in the most similar conditions. Overall, these results question whether LLMs exhibit response consistency.

---

[8] Study 7 also included the GPT-4 model that was used in the comparable study 3. In this case the results were not statistically different over time. However, unlike the other two within-model comparisons, the sample size for study 7 was 100 instead of 1,000, per condition. An unbalanced factorial design may result in problematic estimates, especially when the interaction term is included in the model (Landsheer & van den Wittenboer, 2015).

## General Discussion

The aim of this research is to test the extent to which LLM behavior resembles that of consumers, and more generally humans. In ten pre-registered studies, involving 66,000 conversations, we probe whether different LLMs attenuate or fix, amplify, or reverse human biases, and whether their responses are consistent. Our results cast doubt on the degree to which LLM responses can be assumed to be human-like. In none of the conditions in our 10 studies did we observe effects similar in magnitude and direction to those established for consumers, and more generally humans. Instead, in all 10 studies, we observed bias attenuation, in six studies bias amplification, and in five studies bias reversal, at least for some conditions. Moreover, LLM responses appeared to be inconsistent across models of the same family, across different researchers employing similar methodologies, when handling minor prompt variations, and even when the exact same model responded to the exact same prompt, only with a time delay. We now discuss the implications of these findings.

*Theoretical implications*

Our research has important implications for theories aimed at understanding AI and its usage. First, we add to the research challenging the assumption that LLMs are human-like. Although such human-likeness was almost taken for granted in AI development (Li et al., 2024; Meng, 2024; OpenAI, 2023) and was supported by early evidence (Suri et al., 2024), subsequent investigations show that it may be happening only for about 50% of the cases (Chen et al., 2025). Our work contributes to this development by (a) defining four distinct types of deviations from human-likeness and (b) showing that LLM behavior is much less human-like when we account for those deviation types: We did not find a human-like pattern of responses in any of our studies.

This degree of deviation from human-likeness raises a second point about LLM behavior. Specifically, if we did not observe human-like behavior in any of the 10 studies about well-documented consumers, and more generally human biases and heuristics, then perhaps deviations of LLM behavior from human-likeness may be even more widespread in domains that are not so established, and hence harder to study. Current studies of LLMs are either focused on a single dimension (usually task-specific performance; Gilardi et al., 2023; Hatch et al., 2025) or evaluate LLMs dichotomously (e.g., different from humans or not; Suri et al., 2024; Chen et al., 2025). Our research provides evidence that richer evaluations of LLMs, such as the classification we propose, are necessary to fully understand their behavior. Omitting assessments that allow LLMs to make different mistakes than humans, or that incorporate effect magnitude (McShane et al., 2024), can lead to superficial impressions about the extent to which LLMs can emulate, and hence potentially replace, humans.

Third, the current understanding of AI seems to be excessively human-centric, including, for example, suggestions for employing psychological theories to explain AI behavior (Hoffman et al., 2022). Similarly, the source of AI errors is often thought to be either the human-generated dataset based on which algorithms are trained (Morewedge et al., 2023) or the human-designed training process (Latif & Zhai, 2024; Sweeney, 2013). Our demonstration of reverse biases, as well as response inconsistency over time, is hard to reconcile with these accounts; hence, we call for a deeper investigation of the sources of AI biases and for theories that encompass both human-like and AI-unique characteristics. This call is aligned with evidence that LLMs respond differently than humans for risk-taking and cooperation (Mei et al., 2024), as well as for political stance (Rozado, 2024).

Fourth, our findings contribute to the discussions about the ethical use of AI, which may require accurate representations of consumers, and more generally human behavior. Such representativeness is difficult because human responses are diverse and depend on social and cultural influences (Hermann & Puntoni, 2024). Our findings illustrate a second level of complication for this principle: AI behavior may be different from any known human behavior. For example, the endowment effect is well established for individualistic cultures, but attenuated for collectivistic ones (Maddux et al., 2010). Still, in two studies (3 and 7) we found reversals of the endowment effect, which are very rare for humans (3.66% in Sayman & Öncüler's [2005] meta-analysis). Hence, our findings broaden the question from "which groups is AI behavior representative of?" to "is AI behavior representative of human behavior, in general?" Relatedly, besides representativeness, another desirable property of ethical AI use is unbiasedness (Hermann & Puntoni, 2024). Our findings show that these two properties can be mutually exclusive: In the domains of human biases, representative behaviors of an LLM are, inevitably, biased. Hence, our results suggest prioritizing some of the principles of ethical usage of AI, depending on the purpose of this usage (e.g., performance vs. prediction).

Finally, our research indicates that LLM version and time of observation might infiltrate experimental designs as potential alternative explanations. For example, as the chat version of LLMs is frequently updated, any differences observed by research done over time using chat (e.g., Suri et al., 2024) may be contaminated by the influence of a different version. We find such cross-version differences in all our studies. Additionally, research examining different experimental conditions or domains over time cannot lead to certain conclusions about whether observed effects are a function of the experimental design or time itself. Specifically, our work indicates that the exact same "frozen" version of an LLM can give different responses to the exact same prompt after a time delay (e.g., studies 5.1 vs. 5.2 and 6.1 vs. 6.2). To properly make such comparisons, all experimental conditions should be conducted at the same time. More broadly speaking, persevering within-model response inconsistencies cast doubt on how much trust can be placed on LLM inputs for decision-making and research.

*Practical implications*

Our results also hold important practical implications, as they involve real interactions with the LLMs. The responses we received are the responses that the LLMs would give to anyone asking these questions, as the interactions happened in their "native" environment.

At the level of individuals, in 2025, four hundred million people use ChatGPT every week (Rooney, 2025). Many of these consumers have used this LLM for advice (Krügel et al., 2023). Not knowing that LLMs can present reverse biases (e.g., in our study of the endowment effect) or inconsistent responses may have resulted in suboptimal choices for many of these consumers. At the LLM developer level, not knowing the limitations of the newer model, as they are uncovered by our research, prevents their correction in future models.

At the business level, AI's two most prominent potential functions in organizations are substituting (Li et al., 2023) or aiding (Noy & Zhang, 2023) humans. These functions may aim at improving or emulating consumers, and more generally human behavior. For instance, for performance improvement, GPT-4 had a better performance than 85% of human programmers in coding tasks (Hou & Ji, 2025) or in detecting emotions from facial expressions (Kramer, 2025). In this case, differences in the form of performance improvement, such as bias attenuation compared to humans or older models, are desirable; newer models do better, so they fulfill their purpose of replacing or aiding humans more productively. Performance deterioration, such as bias amplification, on the other hand, is undesirable.

The second aim can be emulating consumers, and more generally human behavior and preferences. For example, when AI is used to replace market research participants (Li et al., 2024), it should mimic their responses and preferences, without trying to improve even the suboptimal ones. In this case, all deviations, even those that improve decision-making and performance, are problematic. Moreover, if AI aims to mimic human behavior, and human behavior is relatively stable (Epstein, 1979), then changes in AI behavior over time likely often prevent the goal of human-likeness. Our results challenge the notion that emulating human behavior can happen "out of the box," without previous testing. When it comes to replacing humans, we show that LLMs often have un-humanlike behavior. When it comes to aiding humans, we show that LLMs tend to exhibit biased behaviors, but in unique ways. Hence, we propose careful planning and testing of AI models before their adoption, especially in industries where accountability is important.

Relatedly, our work is in line with recent research and debunks the myth that the most advanced LLMs necessarily perform better in terms of human-likeness (McShane et al., 2025). Instead, we show systematic reverse biases for GPT-4, the most advanced LLM at the time of our first data collection, or GPT-4o-mini (one of the most advanced models) at the time of our third wave of data gathering. Identifying distinct ways in which LLMs can deviate from human-likeness can help restore or make models more human-like.

Another suggested use of LLMs is to debias consumer decision-making through decision delegation (Meng, 2024). Our evidence suggests caution for this practice, even when LLMs are asked to consider and avoid these biases. In several studies, we have observed bias attenuation, but we have also observed reverse bias or bias amplification. Awareness of these different response tendencies is necessary for the productive use of AI as decision-making support.

At the regulatory level, our results suggest higher scrutiny of the use of LLMs. For example, we find that GPT-3.5 tends to be overwhelmingly risk-seeking. Since this model has been used billions of times, there could have been consequences for those who use it for advice in domains involving risky decisions, such as investments or medical treatments. We propose that each LLM should be made available with a detailed list of its evidence-based characteristics.

*Potential underlying mechanisms*

Understanding why LLMs deviate from human-likeness in distinct ways is an arduous task, beyond the scope of a single research paper. LLMs utilize Deep Learning, a special category of AI, famous for its performance in image, text, and speech recognition, and infamous for its lack of explainability (Goodfellow et al., 2016). In addition, the heuristics and biases we study involve complex processes that have been the focus of scientific inquiry for decades.

The prevalent explanation for LLMs' biases is biased training datasets (Morewedge et al., 2023; Caliskan et al., 2017), which reflect the biases of the human brain black-box (Bonezzi et al., 2022). Although this may still be the driver of some of our results (e.g., in the cases where LLMs replicate directionally human biases), it is probably not the only one. For example, the reverse biases that we found for both models speak against this possibility.

A second source could be LLM training. Does providing more human training data increase LLMs' human-likeness, and hence their degree of human bias, or does it allow LLMs to overcome consumers, and more generally human biases? The answer is not clear. Studies 3, 6.2, and 6.3 showed cases where the better-trained GPT-4 produced reverse biases but the less-trained GPT-3.5 did not. In addition, in study 7, we did not find evidence that better-trained models mimicked human biases more closely, even if they were asked to behave as research participants (see MDA for details). These findings suggest that the training process also cannot be the sole explanation.

A third possibility can be that model advancement also changes the way LLMs prioritize their relative reliance on the provided prompt versus their training data. For instance, more advanced models, like GPT-4 in most of our studies, may rely relatively more on their training base, at the expense of relying on the provided prompts. This could explain why GPT-4, but not GPT-3.5, amplified the representativeness and the availability heuristics. For example, the representativeness heuristic might be explained by an over-reliance on training data, including numerous instances that align with the "representative" description, and an under-reliance on the prompt, a good understanding of which would hint towards the correct answer.

*Limitations and future research*

We acknowledge the limitations of our work and propose ways that future research can overcome them. First, it could be argued that the results are contingent on the specific prompts we used. We have tried to overcome this limitation in two ways. First, our prompts closely followed established demonstrations of the effects we studied. Second, we did systematically vary these prompts across studies, varying prompts for debiasing and role assignment. Of course, with the right prompts, deviations from human likeness may be attenuated or completely disappear. Optimally prompting, though, is a separate issue from the chronic capabilities and characteristics of a model (Henrickson & Meroño-Peñuela, 2023). We encourage future research that systematically varies prompts to identify boundary conditions for our effects.

Relatedly, as described above, although the LLMs had access to the results of the original studies we benchmarked this research on, they did not closely mimic these results in any study, even when asked to take the role of research participants. The reasons behind this lack of human-likeness are puzzling, and we leave it to future research. On top of this, conditions under which each distinct type of deviation from human-likeness is more likely, and ways to improve on each of these deviations, provide fruitful future research directions. This is mostly true for the case of reverse biases, which are hard to explain by sources such as training datasets and processes.

Third, a potential criticism of studies 3, 5, and 6 is that the LLM was asked to indicate its preference, although it is implausible that it has one (Arcas, 2022). We framed our questions this way to replicate previous research and avoid responses purely based on LLM's training (e.g., "How much is a beer worth?" could induce the LLM to simply reflect the usual prices of beer).

Fourth, our studies were conducted with the API version of the LLMs, which sometimes provides different responses than the Chat version (Hagendorff et al., 2023). This difference may be a result of the constant updating of the Chat version's algorithm, which often incorporates previous discussions with them by default (OpenAI, 2024); this speculation also stems from observing shorter update cycles for the Chat version. We chose the API version because, first, it allowed us to study LLM behavior at scale, and second, it is used daily by the most important companies in the world (Rooney, 2025). We acknowledge that further research should study how and when the two versions of the same LLM may provide divergent responses.

Finally, in most studies, we only tested the human-likeness of GPT-3.5 and GPT-4. We based this decision on their performance and popularity at the time of the studies. Although study 7, involving more state-of-the-art models, supported our conclusions by showing three deviations from human-likeness (bias attenuation, reversal, and inconsistent responses), future research should expand our results in LLMs outside the GPT family.

## Conclusion

In conclusion, this research raises considerations about four different ways in which LLMs may deviate from human-likeness. Evidence from ten studies, spanning six seminal domains in the understanding of consumers, and more generally human psychology, shows that human-likeness cannot be taken for granted. Hence, we call for a more open view about the development and understanding of AI models, and a re-formulation of the key assumptions in the field based on proper empirical testing.

## Data, Materials, and Software Availability

All our study materials (used prompts, datasets, code, and pre-registration forms) can be found at https://osf.io/q5nmw/?view_only=200733a794c44ad3b1811ce9e81249b3

## Acknowledgments

The data for the studies in this paper was collected through OpenAI API using two LLMs. Exact prompts were pre-registered and disclosed in detail in the MDA and at

https://osf.io/q5nmw/?view_only=200733a794c44ad3b1811ce9e81249b3

The authors declare no competing interest, financial or non-financial. Specifically, the authors have no personal relationships, professional affiliations, advisory roles, board memberships, or other non-financial interests that could be influencing the work reported in this manuscript. No third party had the right to review or require changes to the paper prior to submission.

This work was funded by:

MCIN /AEI /FEDER, UE Grant No. PID2022-138729OA-I00

MCIN /AEI /FEDER, UE Grant No. PID2020-119622GA-I00

MCIN /AEI /FEDER, UE Grant No. PID2022-139380NA-I00